\documentstyle [12pt]{article}
\newcommand{\be}{\begin{equation}}
\newcommand{\ee}{\end{equation}}
\newcommand{\ber}{\begin{eqnarray}}
\newcommand{\eer}{\end{eqnarray}}
\newcommand{\bers}{\begin{eqnarray*}}
\newcommand{\eers}{\end{eqnarray*}}

\begin{document}
\rightline{\vbox{\baselineskip=14pt
\rightline{UH-511-942-99} \break
  \rightline{August 1999}
}}
\def\ee{\end{equation}}
\def\ra{\rightarrow}
\thispagestyle{empty}

\vspace {.5in}
\begin{center}
{\Large \bf{CP Violation in Hyperon Decays\footnote{Invited talk
presented at Kaon '99, Chicago, June 21-26, 1999.} \\}}
\vspace{.5in}
{\rm {Sandip Pakvasa} \\}

\vskip .3in

{\it Department of Physics and Astronomy \\}
{\it  University of Hawaii \\}
{\it Honolulu, Hawaii 96822\\}
\vskip .5in
\end{center}  
\begin{abstract}
The theory and phenomenology of CP violation in hyperon
decays is summarized.
 \end {abstract}
\newpage
\baselineskip 24pt

\section{Introduction}

The CPT theorem was proved in 1955\cite{luders} and soon thereafter 
L$\ddot{\rm{u}}$ders and 
Zumino\cite{luderss} deduced from it the equality of masses and lifetimes 
between particles and
anti-particles. In 1958 Okubo\cite{okubo} observed that CP violation allows 
hyperons and
antihyperons to have different branching ratios into conjugate channels
even though their total rates must be equal by CPT. Somewhat later, this paper
inspired Sakharov\cite{sakha} to his famous work on cosmological baryon-antibaryon
asymmetry.  Pais\cite{pais} extended Okubo's proposal to asymmetry parameters in
$\Lambda$ and $\bar{\Lambda}$ decays. The subject was revived in the
'80s and a number of calculations were made\cite{brown,chau}.  Only now, 
over 40 years after Okubo's paper, are 
these proposals about to be tested in the laboratory.

The reason for the current interest is the need to find CP violation in
places other than just $K_L-K_S$ complex.  Only a number of different
observations of CP violation in different channels will help us pin down
the source and nature of CP violation in or beyond the standard model (SM).
From this point of view hyperon decay is one more weapon in our arsenal
in addition to
the K system, the B system, the D system,  etc.

\section{Phenomenology of Hyperon Decays}

I summarize here the salient features of the phenomenology of
non-leptonic hyperon decays \cite{rosen}.  Leaving out $\Omega^-$ decays, there  are
seven decay modes $\Lambda \ra N \pi, \ \Sigma^\pm \ra N \pi$ and $\Xi \ra \Lambda \pi$.  The
effective matrix element can be written as 
\begin{equation}
i \ \bar{u}_{\bar{p}} (a + b \gamma_5) u_\Lambda \ \phi
\end{equation}
for the mode $\Lambda \ra p + \pi^-$, where a and b are complex in general.
The corresponding element for $\bar{\Lambda} \ra \bar{p} + \pi^+$ is
then:
\begin{equation}
i \ \bar{v}_{\bar{p}} (-a^* + b^* \gamma_5) v_{\bar{\Lambda}} \phi^+
\end{equation}
It is convenient to express the observables in terms of S and  P and
write the matrix element as
\begin{equation}
S + P \ {\bf \sigma}.\hat{\bf q}
\end{equation}
where {\bf q} is the proton momentum in the $\Lambda$ rest frame
and S and P are:
\begin{eqnarray}
S &=& a \sqrt{ \left\{ (m_\Lambda + m_p)^2 - m^2_\pi \right\}
\over 
16 \pi \ m^2_{\Lambda}} \nonumber \\
P &=& b \sqrt {\left\{ (m_\Lambda - m_p)^2 - m^2_\pi \right\}
\over 16 \pi \ m^2_{\Lambda}}
\end{eqnarray}
In the $\Lambda$ rest-frame, the decay distribution is given by:
\begin{eqnarray}
\frac{d \Gamma}{d \Omega} &=&
\frac{\Gamma}{8 \pi} %
\{ 
[1 + \alpha < {\bf \sigma}_\Lambda > . \hat{\bf \sigma} ] \nonumber \\
&+& < {\bf \sigma}_p >. [( \alpha + < {\bf \sigma}_\Lambda > . 
\hat{\bf q}) \hat{\bf q}+ \ \ \beta < {\bf \sigma}_\Lambda > \times \hat{\bf q} \nonumber \\
&+& \gamma ( \hat{\bf q} \times ( < {\bf \sigma}_\Lambda > \times
 \hat{\bf q})] 
 \}
\end{eqnarray}  
where $\Gamma$ is the decay rate and is given by:
\begin{equation}
\Gamma = 2 \mid {\bf q} \mid \left\{ \mid S \mid^2 + \mid P \mid^2 \right \}
\end{equation}
$\alpha, \beta$ and $\gamma$ are given by
\begin{eqnarray}
\alpha &=& 2Re (S^*P ) \over \left \{ \mid S \mid^2 + \mid P \mid^2 \right
\}, \nonumber \\
\beta &=& 2 Im (SP^*) \over \left \{ \mid S \mid^2 + \mid P \mid^2
\right \}
\nonumber \\
\gamma &=&  \left \{ \mid S \mid^2 - \mid P \mid^2 \right \} \over 
\left \{ \mid S \mid^2 + \mid P \mid^2 \right \}
\end{eqnarray}
For a polarized $\Lambda$, the up-down asymmetry of the final proton is
given by  $\alpha ( \alpha$ is also the longitudinal polarization of the
proton for an unpolarized $\Lambda)$.  $\beta$ and $\gamma$ are
components of the transverse polarization of proton \cite{lee}.

The observed properties of the hyperon decays can be summarised as:  (i)
the $\Delta I = 1/2$ dominance i.e. the $\Delta I = 3/2$ amplitudes are about
5\% of the $\Delta I = 1/2$ amplitudes; (ii) the asymmetry parameter
$\alpha$ is large for $\Lambda$ decays, $\Xi$ decays and $\Sigma^+ \ra p
\pi^0$ and is near zero for $\Sigma^\pm \ra n \pi^\pm$; and (iii) the
Sugawara-Lee triangle sum rule $\sqrt{3}A ( \Sigma^+ \ra p  \pi^0) - A 
(\Lambda \ra p \pi^-) = 2A (\Xi \ra \Lambda \pi^-)$ is satisfied to a level
of 5\% in both $s$ and $p$ wave amplitudes.

\section{CP Violating Observables}

Let a particle P decay into several final states $f_1, f_2$ etc.  The
amplitude for P $\ra f_1$ is in general:
\begin{equation}
A= A_1 e^{i \delta 1} \ \ + A_2 \ e^{i \delta 2}
\end{equation}
where 1 and 2 are strong interaction eigenstates and $\delta_i$ are
corresponding final state phases.  Then the amplitude for
$\bar{P} \ra \bar{f}_1$ is
\begin{equation}
\bar{A} = A^*_1 e^{i \delta 1} \ \ + A^*_2 \ e^{i \delta 2}
\end{equation}
If $\mid A_1 \mid >> \mid A_2 \mid$, then the rate asymmetry
$\Delta ( = (\Gamma - \bar{\Gamma}) / (\Gamma + \bar{\Gamma}) )$ is
given by:
\begin{equation}
\Delta \approx -2 \mid A_2/A_1 \mid sin ( \phi_1 - \phi_2) sin(\delta_1- 
\delta_2)
\end{equation}
where $A_i = \mid A_i \mid e^{i \phi_i}$.  Hence, to get a non-zero rate
asymmetry, one must have (i) at least two channels in the final state,
(ii) CPV weak phases must be different in the two channels, and (iii) 
\underline{unequal} final state scattering phase shifts in the two
channels\cite{brown}. 
A similar calculation of the asymmetry of $\alpha$\cite{overseth} shows that
for a single isospin channel dominance,
\begin{equation}
A= (\alpha + \bar{\alpha})/ (\alpha - \bar{\alpha}) = 2 tan ( \delta_s -
\delta_p) \ sin (\phi_s - \phi_p)
\end{equation}
In this case the two channels are orbital angular momentum $0$ and $1$;
and even a single isospin mode such as $\Xi^- \ra \Lambda \pi^-$ can
exhibit a non-zero A. In B decays an example of a single isospin mode
exhibiting CP violating rate asymmetry is $B \ra \pi \pi$, i.e. 
In this case the two eigen-channels with different weak CP phases and
different final state phases are $B \ra D \overline{D} \pi \ra \pi\pi$
and $B \ra \pi \pi \ra \pi \pi$\cite{wolfenstein}. 

To define the complete set of CP violating observables, consider the
example of the decay modes $\Lambda \ra p \pi^-$ and
$\bar{\Lambda} \ra \bar{p} \pi^+$.  The amplitudes are:
\begin{eqnarray}
S&=& - \sqrt{2 \over 3} S_1 e^{i ( \delta_1 + \phi_1^s)} +
    \frac{1} {\sqrt{3}} S_3 e^{i ( \delta_3 + \phi^s_3)} \nonumber \\
P&=& - \sqrt{2 \over 3} P_1 e^{i ( \delta_{11} + \phi_1^p)} +
    \frac{1}{\sqrt{3}} P_3 e^{i ( \delta_3 + \phi^p_3)} 
\end{eqnarray}
where $S_i, P_i$ are real, $i$ refers to the final state isospin (i=2I)
and $\phi_i$ are the CPV phases.  With the knowledge that $S_3/S_1$,
$P_3/P_1 <<$ 1 ;  one can write\cite{donog,donog1}
\begin{eqnarray}
\Delta_\Lambda &=& \frac{(\Gamma-\overline{\Gamma})}{(\Gamma + \overline{\Gamma})}
\cong \sqrt{2} \ (S_3/S_1) sin ( \delta_3 - \delta_1) sin
(\phi_3^s - \phi_1^s) \nonumber \\
A_\Lambda &=& \frac{(\alpha +\overline{\alpha})} {(\alpha - \overline{\alpha})}
\cong - tan (\delta_{11} - \delta_1) sin (\phi_1^p - \phi_1^s)
\nonumber \\
B_\Lambda &=& \frac{(\beta +\overline{\beta})}{(\beta - \overline{\beta})}
\cong  cot  (\delta_{11} - \delta_1) sin (\phi_1^p - \phi_1^s)
\end{eqnarray}
For $\pi$N final states, the phase shifts at $E_{c.m.} = m_\Lambda$ are
known and are: $\delta_1 = 6^0, \ \delta_3 = -3.8^0, \ \delta_{11} = 1.1^0$
and $\delta_{31} = -0.7^0$.  The CPV phases $\phi_i$ have to be provided
by theory.

Similar expressions can be written for other hyperon decays.  For
example, for $\Lambda \ra n \pi^0$, $\Delta$ is $- 2 \Delta_\Lambda$
and $A$ and $B$ are identical to $A_\Lambda$ and $B_\Lambda$.  For
$\Xi^- \ra \Lambda \pi^-$ ( and $\Xi^0 \ra \Lambda \pi^0)$ the 
asymmetries are\cite{donog1}:
\begin{eqnarray}
\Delta_\Xi & = & 0    \nonumber \\
A_\Xi & = &-tan (\delta_{21} - \delta_2) sin (\phi^p - \phi^s)
\nonumber \\
B_\Xi & = & cot (\delta_{21} - \delta_2) sin (\phi^p- \phi^s)
\end{eqnarray}
where $\delta_{21}$ and $\delta_2$ are the $p$ and $s$-wave $\Lambda
\pi$ phase shifts at $m_\Xi$ respectively.  Somewhat more
complicated expressions can be and have been written for $\Sigma$ decays\cite{donog1}.
There are no  experimental proposals at the moment for measuring these.

\section{Calculating CP Phases}

In standard model description of the non-leptonic hyperon decays, the
effective $\Delta S = 1$ Hamiltonian is
\begin{equation}
H_{eff} = \frac{G_F}{\sqrt{2}} \ U_{ud}^* \ U_{us} \sum^{12}_{i=1} c_i
(\mu) \ O_i ( \mu)
\end{equation}
after the short distance QCD corrections  (LLO + NLLO) where
$c_i = z_i + y_i \tau ( \tau = -U_{td} \ U^*_{ts} / U_{ud} \ U_{us})$,
and $\mu \sim 0(1$ GeV). 
For CP violation, the most important operator is:
\begin{equation}
O_6 = \bar{d} \ \lambda_i\gamma_\mu (1+ \gamma_5) s 
\bar{q} \lambda_i \gamma_\mu (1 -\gamma_5)q
\end{equation}
and $y_6 \approx -0.1$ at $\mu \sim 1 \ GeV$.  
To estimate the CP phases in Eq. (12), one adopts the following
procedure. The real parts(in the approximation that the imaginary 
parts are very small) are known from the data on rates and asymmetries.
The real parts of the amplitudes have also been evaluated in SM with
reasonable success with some use of chiral perturbation theory(current
algebra and soft pion theorems) and a variety of choices for the 
baryonic wave functions. The MIT bag model wave function is one such
choice which gives conservative results.
The same procedure is adopted for calculating the imaginary parts using 
$O_6$. The major 
uncertainty is in the hadronic matrix elements and the fact that the 
simultaneous fit of S and P waves leaves a factor of 2 ambiguity
\cite{donog2}.
In the SM,
with the Kobayashi-Maskawa phase convention there is no CPV in $\Delta I
= 3/2$ amplitudes; and for $\Lambda$ decays $\phi_3 = 0$. There is a small
electroweak penguin contribution to $\phi_3$ which is neglected. The
rate asymmetry is dominated by the s wave amplitudes and the asymmetry $A_\Lambda$
is dominated by the $\Delta I = 1/2$ amplitudes. Evaluating the 
matrix elements in the standard way and with the current knowledge of the 
K-M matrix elements one finds for the decays\cite{donog1,iqbal} $\Lambda \ra p \pi^-$ and $\Xi^-
 \ra \Lambda \pi^-$:
\begin{eqnarray}
& &\phi^s_\Lambda - \phi^p_\Lambda \cong 3.5.10^{-4}  \nonumber \\
& &\phi_\Xi^s - \phi^p_\Xi \cong - 1.4.10^{-4}
\end{eqnarray}
With the $N \pi$  phase shifts known to be 
\begin{equation}
\delta_s - \delta_p \cong 7^0
\end{equation}
one finds for the asymmetry $A_\Lambda$ in the standard model a value of 
about $-4.10^{-5}$.   For the $\Xi \ra \Lambda \pi^-$ decay mode the phase 
shifts are not known experimentally and have to be determined 
theoretically.  There are calculations from 1965 \cite{martin} which gave large values for 
$\delta_s - \delta_p \sim -20^0$; however, all recent calculations based on chiral 
perturbation theory, heavy baryon approximation etc. agree that $\delta_s - 
\delta_p$ lies between $1^0$ and $3^0$ \cite{lu}.  In this case the asymmetry 
$A_\Xi$ is expected to be $\sim - (0.2$ to $0.7) 10^{-5}$. In the Table 1,
the SM results for the expected asymmetries in SM are given. Using very crude
back of the envelope estimates, similar results are obtained.  What is
needed is some attention to these matrix elements from the Lattice community. 

An experimental measurement of the phase shifts $\delta_s - \delta_p$ in 
$\Lambda \pi$ system will put the predictions for $A_\Xi$ on a firmer basis.  
There is an old proposal due to Pais and Treiman \cite{pais1} to measure $\Lambda \pi$ 
phase shifts in $\Xi \ra \Lambda \pi e\nu$, but this does not seem
practical in the near future.  Another technique, more feasible, it to measure $\beta$ and 
$\alpha$ to high precision in $\Xi$ and $\overline{\Xi}$ decays.  Then the 
combination.
\begin{equation}
( \beta - \bar{\beta}) / (\alpha - \bar{\alpha}) = tan \ (\delta_s -
\delta_p)
\end{equation}
can be used to extract $\delta_s -\delta_p$.  To the extend CP phases are 
negligible one can also use the approximate relation:
\begin{equation}
\beta/\alpha \approx tan (\delta_s -\delta_p)
\end{equation}

\section{Beyond Standard Model}

Can new physics scenarios in which the source of CP violation is not K-M 
matrix yield large enhancements of these asymmetries?   We consider some 
classes of models where these asymmetries can be estimated more or less
reliably 
\cite{donog,donog1}.

First there is the class of models which are effectively super-weak \cite{some}.
Examples are models in which the K-M matrix is real and the observed CP
violation is due to exchange of heavier particles; heavy scalars with FCNC,
heavy quarks etc.  In all such models direct CP violation is negligible and
unobservable and so all asymmetries in hyperon decays are essentially zero.
Furthermore, they need to be modified to accommodate the fact that
direct CP violation has now been seen in the kaon decays( the fact that
$\epsilon'/\epsilon$ is not zero).
In the three Higgs doublet model with flavor conservation imposed, the 
charged Higgs exchange tend to give large effects in direct CP violation as 
well as large neutron electric dipole moment \cite{wein}.

There are  two generic classes of left-right symmetric models:  (i) Manifest Left-
Right symmetric model without $W_L - W_R$ mixing \cite{mohap} and (ii) with $W_L -
W_R$ mixing \cite{chang}.  In (i) $U_{KM}^L =$ real and $U_{KM}^R$ complex 
with arbitrary phases but angles given by $U_{KM}^L$.  Then one gets 
the ``isoconjugate'' version in which
\begin{equation}
H_{eff} = \frac{G_F \ U_{us}} {\sqrt{2}}
\left [ J^\dagger_{\mu L} \ J_{\mu L} + \ \eta e^{i \beta} J^\dagger_{\mu R} \ J_{\mu
R}
\right ]
\end{equation}
where $\eta = m^2_{WL} /m^2_{WR}$ and $\beta$ is the relevant CPV phase.  Then
$H_{p.c.}$ and $H_{p.v.}$ have overall phases $(1 + i \eta \beta)$ and
$(1-i \eta  \beta)$ respectively.  To account for the observed CPV in
K-decay, $\eta \beta$
has to be of order $4.5.10^{-5}$.  In this model, $\epsilon'/\epsilon = 0$
and there are no rate asymmetries in hyperon decays but the asymmetries
 A and B are not zero and e.g. A goes as
$2 \eta \beta \sin (\delta_s - \delta_p)$.  In the class of models where
$W_L - W_R$ mixing is allowed, the asymmetries can be enhanced, and also
$\epsilon'/\epsilon$ is not zero in general.

Models where the gluon dipole operator 
is enhanced beyond the SM value are especially interesting: 
\begin{equation}
\overline{d} \lambda^\alpha (a + b \gamma_5) s \ G_{\mu_\nu}^\alpha
\end{equation}
In
SM, the coefficient of this operator is too small to be interesting.
The parameter $a$(actually its imaginary part) is constrained
by the known value of $\epsilon$ and contributes only to hyperon
decays. 
The parameter $b$ can contribute to
both $\epsilon'$ as well as the CP violating asymmetries in hyperon
decays. The current range of $\epsilon'$ as given by the experimental
values we heard here\cite{22} allows {\it b} to contribute to $A_\Lambda$ at a 
level of $5.10^{-4}$\cite{he}. Such enhancement of this operator takes place 
naturally in models where CP violation occurs due to the exchange
of charged higgs scalars(such as the Weinberg model) and can also
occur in several scenarios based on supersymmetry \cite{masiero}.
\begin{table}
\caption{Expectations for Hyperon CPV Asymetries.}
\begin{center}
\begin{tabular}{|lccccc|} \hline
                   &   \mbox{SM}   & \mbox{2-Higgs}    &   \mbox{FCNC}  &  L-R-S  &  L-R-S  \\ 
                   &               &    	       &  \mbox{Superweak}  & (1)   &  (2) \\ 
$\Delta_\Lambda$   &  $10^{-6}$    &   $10^{-5}$       &        0
                                                                            &   0    &  0  \\
$A_\Lambda$   &  $ -4.10^{-5}$    &   $-2.10^{-5}$  &      0  & $-10^5$ &   $6.10^{-4}$  \\
$B_\Lambda$   &$10^{-4}$             &   $2.10^{-3}$     &  0    &
                   $7.10^{-4}$  & -  \\
$\Delta_\Xi$   & 0  & 0   &  0   &  0  &  0  \\
$A_\Xi$   &  $-4.10^{-6}$    &   $-3.10^{-4}$  & 0    & $2.10^{-5}$     & $10^{-4}$  \\
$B_\Xi$   &  $10^{-3}$    &   $ 4.10^{-3}$  & 0    & $3.10^{-4}$     &   -
                   \\ \hline
\end{tabular}
\end{center}
\end{table}

\section{Experiments}

There have been several proposals to detect  hyperon decay asymmetries in
$\bar{p} p \ra \bar{\Lambda} \Lambda, \ \bar{p}{p} \ra \overline{\Xi} \Xi$ and in
$e^+e^- \ra J/\psi \ra \Lambda \overline{\Lambda}$ but none of these
were approved \cite{hamman}.  The only approved and on-going experiment is E871 at
Fermilab.  In this experiment $\Xi^-$ and $\overline{\Xi}^+$ are produced
and the angular distribution of $\Xi^-  \ra \Lambda \pi^- \ra p \pi^- \pi^-$ and
$\overline{\Xi}^+$ compared.  This experiment effectively measures $A_\Lambda +
A_\Xi$ and will be described in detail by Sharon White\cite{white}.
To summarize the implications for the measurement of $A_\Lambda + A_\Xi$
by E871: the SM expectation is about $-4.10^{-5}$ with a factor of two
uncertainty; if new physics should contribute it could be as large as 
$7.10^{-4}$. A measurement by E871 at the $10^{-4}$ level, therefore, will already 
be a strong discriminant.

\section{$\epsilon'/\epsilon$ and Hyperon Decay Asymmetries}

It might seem that now that $\epsilon'/\epsilon$ has been measured and
direct CP violation in $\Delta S=1$ channel been observed, a study of CP
violation in hyperon decays is unnecessary and no new information will
be obtained. Why is it worthwhile measuring
another $\Delta S=1$ process like hyperon decay?  The point is that there
are important differences and the two are not at all identical.
First, there are important differences in the matrix elements. Hyperon
matrix elements do not have the kind of large cancellations that plague
the kaon matrix elements. The hadronic uncertainties are present for both,
but are different. Next, a very important difference is the fact
that the K $\ra \pi \pi$ decay (and hence $\epsilon'$) is only sensitive
to CP violation in the parity violating amplitude and cannot yield any
information on parity conserving amplitudes. Hyperon decays, by contrast, are
sensitive to both. Thus, $\epsilon'/\epsilon$ and hyperon decay
CP asymmetries are different and complimentary. The hyperon decay
measurements are as important and significant as $\epsilon'/\epsilon$.

\section*{Conclusion}

The searches for direct CPV are being pursued in many channels.  $K \ra 2 
\pi, \Lambda \ra N {\pi}$, B decays and D decays.  Any observation of
a signal would be the first outside of $K^0 - \overline{K}^0$ system and
would be complimentary to the measurement of $\epsilon'/\epsilon$. 
This will constitute one more step in our bid to
confirm or demolish the Standard Kobayashi-Maskawa description of CP 
violation.

Hyperon decays offer a rich variety of CP violating observables, each
with different sensitivity to various sources of CP violation.  For
example, $\Delta_\Lambda$ is mostly sensitive to parity
violating amplitudes, $\Delta_{\Sigma +}$ is sensitive only
to parity conserving amplitudes, $A$ is sensitive to both etc.  The
number of experimental proposals is rather small so far.  The one
on-going experiment Fermilab E871 can probe $A$ to a level of $10^{-4}$
which is already in an interesting range.  In addition to more
experiments, this subject sorely needs more
attention devoted to calculating the matrix elements more reliably.
   
\section*{Acknowledgment}

I am grateful to my collaborators Alakabha Datta,
Xiao-Gang He, Pat O'Donnell, German Valencia and John Donoghue and to
members of the E871 collaboration for many discussions. 
The hospitality of Jon Rosner, Bruce Winstein and
colleagues was memorable and the atmosphere of the conference was 
most stimulating. This work is supported in part by USDOE under Grant \#DE-
FG-03-94ER40833.


\end{document}